\documentclass[preprint2, times, twocolappendix]{aastex631}

\newcommand{\mast}{\,\text{mag}\,\text{arcsec}^{-2}}

\newcommand{\projecttitle}{Gnuastro: visualizing the full dynamic range in color images}
\newcommand{\projectversion}{48f5408}
\newcommand{\projectgitrepo}{https://codeberg.org/gnuastro/papers}
\newcommand{\projectgitbranch}{color-faint-gray}
\newcommand{\projectcopyrightowner}{Infante Sainz Raul <infantesainz@gmail.com>}
\newcommand{\gnuastroversion}{0.21.43-3101}

\newcommand{\maneageversion}{8161194}

\newcommand{\objname}{M51}

\newcommand{\chr}{iSDSS}
\newcommand{\chg}{rSDSS}
\newcommand{\chb}{gSDSS}
\newcommand{\chh}{J0660}

\newcommand{\sbch}{rSDSS}
\newcommand{\sbfaint}{25}
\newcommand{\sbbright}{23}
\newcommand{\jplusdr}{3}

\ifdefined\highlightnew

\else

\fi

\ifdefined\highlightnotes
\newcommand{\tonote}[1]{\textcolor{red!60!black}{[#1]}}
\else
\newcommand{\tonote}[1]{{}}
\fi

\usepackage{graphicx}

\usepackage{xcolor}
\color{black} % Color of main text.
\definecolor{DarkBlue}{RGB}{0,0,90}

\hypersetup{
    pdftitle={\projecttitle},
    pdfauthor={\projectcopyrightowner},
    pdfsubject={\projectgitrepo{} (commit \projectversion)},
    pdfkeywords={Gnuastro, Maneage, RGB, color, faint, gray, J-PLUS}
}

\usepackage{tikz}
\usetikzlibrary{external}
\tikzsetexternalprefix{tex/tikz/}

\usepackage{pgfplots}
\pgfplotsset{compat=newest}
\usepgfplotslibrary{groupplots}
\pgfplotsset{
  axis line style={thick},
  tick style={semithick},
  tick label style = {font=\footnotesize},
  every axis label = {font=\footnotesize},
  legend style = {font=\footnotesize},
  label style = {font=\footnotesize}
  }

\shorttitle{\projecttitle}
\shortauthors{Infante-Sainz, R.}

\newcommand{\addimage}[6]{

  \node[anchor=south west] (img) at (#1\linewidth,#2\linewidth)
       {\includegraphics[width=0.49\linewidth]
         {tex/build/figures/#3.pdf}};

  \node[fill=white, minimum width=0.01\linewidth, anchor=south west]
   at (#4\linewidth,#5\linewidth)
       {\textcolor{black}{#6}};

}

\begin{document}

%% Title
\title{\projecttitle}

%% Authors
\author[0000-0002-6220-7133]{Ra\'ul Infante-Sainz}
\author[0000-0003-1710-6613]{Mohammad Akhlaghi}
\affiliation{Centro de Estudios de F\'isica del Cosmos de Arag\'on (CEFCA), Plaza San Juan, 1, E-44001, Teruel, Spain}

\correspondingauthor{Ra\'ul Infante-Sainz}
\email{infantesainz@gmail.com}

%% Abstract
\begin{abstract}
  \noindent
  Color plays a crucial role in the visualization, interpretation, and analysis of multi-wavelength astronomical images.
  However, generating color images that accurately represent the full dynamic range of astronomical sources is challenging.
  In response, Gnuastro v0.22 introduces the program \texttt{astscript-color-faint-gray}, which is extensively documented in the Gnuastro manual.
  It employs a non-linear transformation to assign an 8-bit RGB (Red-Green-Blue) value to brighter pixels, while the fainter ones are shown in an inverse grayscale.
  This approach enables the simultaneous visualization of low surface brightness features within the same image.
  This research note is reproducible with Maneage, on the Git commit \projectversion.
\end{abstract}

%% Keywords (from https://astrothesaurus.org)
\keywords{Astronomy software (1855), Open source software (1866), Astronomy data visualization (1968), Low surface brightness galaxies (940)}

%% Start of main body.
\section{Introduction}
\label{sec:introduction}
\noindent
In common language, an image is a visual representation of something.
More technically, an image is an array of numerical values or pixels (picture elements).
High pixel values mean more intensity or flux, while low pixel values imply lower brightness.

Astronomical images have a very wide range of pixel values.
On the one hand, there are regions with very high signal-to-noise ratios (S/N$>$1000) which correspond to the brightest parts in the center of stars or galaxies.
On the other hand, there are regions for which there is no significant signal from astronomical sources at all: the noisy background pixels.
Intermediate cases are those regions with a per-pixel S/N smaller than three.
These include the outer part of galaxies, stellar streams, diffuse emission like Galactic cirrus, and other faint structures.
Typically, the distribution of pixel values in an astronomical image is highly non-uniform.
The majority of pixel values tend to be faint, while only a few stand out as exceptionally bright.
Combined with the 8-bit (256 layers) nature of standard image formats (like JPEG or PNG; as opposed to 32-bit FITS images), visualization of astronomical images becomes a challenging task.

Astronomical images are usually captured using cameras equipped with filters.
These filters allow specific wavelengths of light to pass through while blocking others, creating what is known as a ``channel''.
A single-channel image does not inherently possess color information since the color is defined by multiple channels, typically the RGB channels (for Red, Green, and Blue).

Color images in common formats serve a more profound purpose than just producing beautiful pictures for outreach purposes.
They contain valuable physical information that can distinguish between astronomical sources.
Variations in color result from differences in stellar populations, ages, distance, and more.
Such color information can be automatically analyzed by specialized software and artificial intelligence algorithms.
Consequently, addressing the challenges of creating high-quality color images is vital for extracting this valuable physical information.

Gnuastro's \texttt{astscript-rgb--faint-gray} is a new executable program to address this problem.
It is introduced in Gnuastro v0.22\footnote{This paper is published before the release of Gnuastro v0.22. If not yet available, please use the latest test (alpha) release.} and addresses the complexities commented above by applying the stretching technique described by \citet{lupton2004} for the brighter pixels, and using inverse-grayscale (where the brightest is black) for the noisy regions.
This combination avoids the ``saturation'' of the extended/diffuse signal in a black background, making it particularly important in the context of very deep imaging and low surface brightness studies \citep[see e.g.,][]{infantesainz2020,trujillo2021,martinezdelgado2023,roman2023}.

\section{Creating color images}
\label{sec:creatingcolorimages}
\noindent

This script takes three input images, typically representing the different RGB channels.
Once the transformations are applied, the modified images are combined to construct the color image.
The script offers various options to customize the resulting color image.
Traditional black background images hide interesting details that are revealed by considering a gray background.
Additionally, it is possible to provide a fourth image (commonly referred to as ``K'') that is used to represent the gray background pixels, resulting in a combination of R, G, B, and K channels.

It is important to note that the specific parameters used in the script may need adjustment based on the characteristics of the input images.
Therefore, the generation of a customized color image might involve some trial and error to achieve the desired result.
For comprehensive details on how to use \texttt{astscript-color-faint-gray}, we refer to the Gnuastro manual\footnote{\url{https://www.gnu.org/software/gnuastro/manual}}.
In particular, there is a step-by-step tutorial on how to use this program with real examples directly executable on the command-line that is always up-to-date with the latest version of the software.

\begin{figure*}[t]
  \ifdefined\makepdf%
    \tikzsetnextfilename{fig-labeling}%
    \begin{center}
\begin{tikzpicture}

 \addimage{0.0}{0.0}{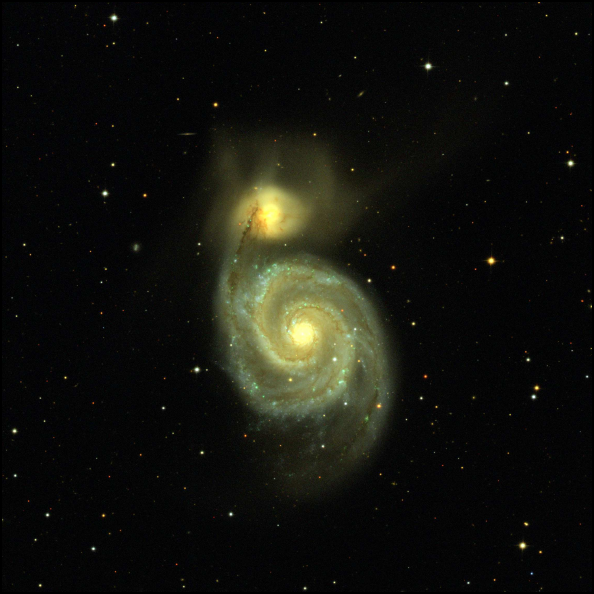}{0.02}{0.45}{R-G-B only color}

 \addimage{0.5}{0.0}{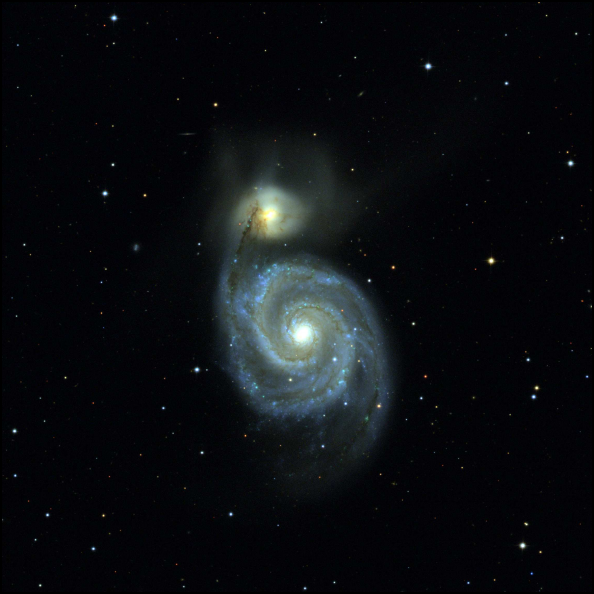}{0.52}{0.45}{R-G-B only color (weighted)}

 \addimage{0.0}{-0.5}{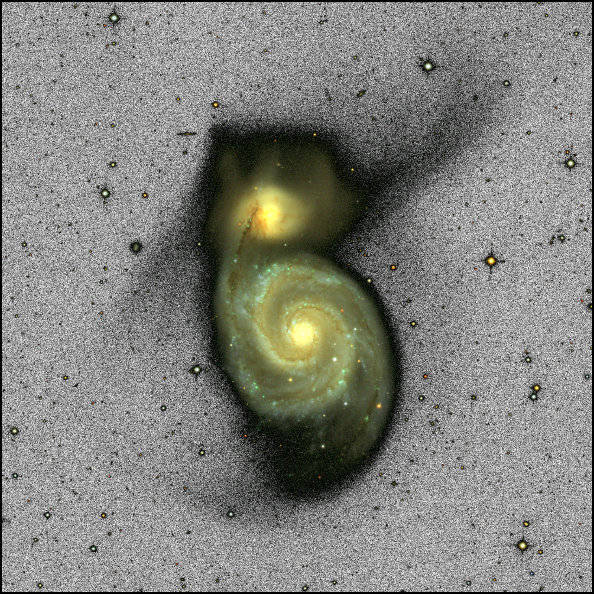}{0.02}{-0.05}{R-G-B grayscale for faint}

 \addimage{0.5}{-0.5}{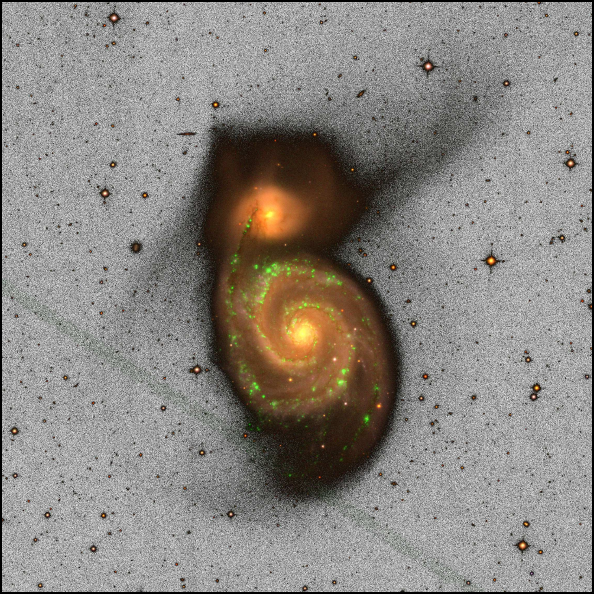}{0.52}{-0.05}{R-H$\alpha$-B grayscale for faint}

\end{tikzpicture}
\end{center}%
  \else
    \includegraphics[width=\linewidth]{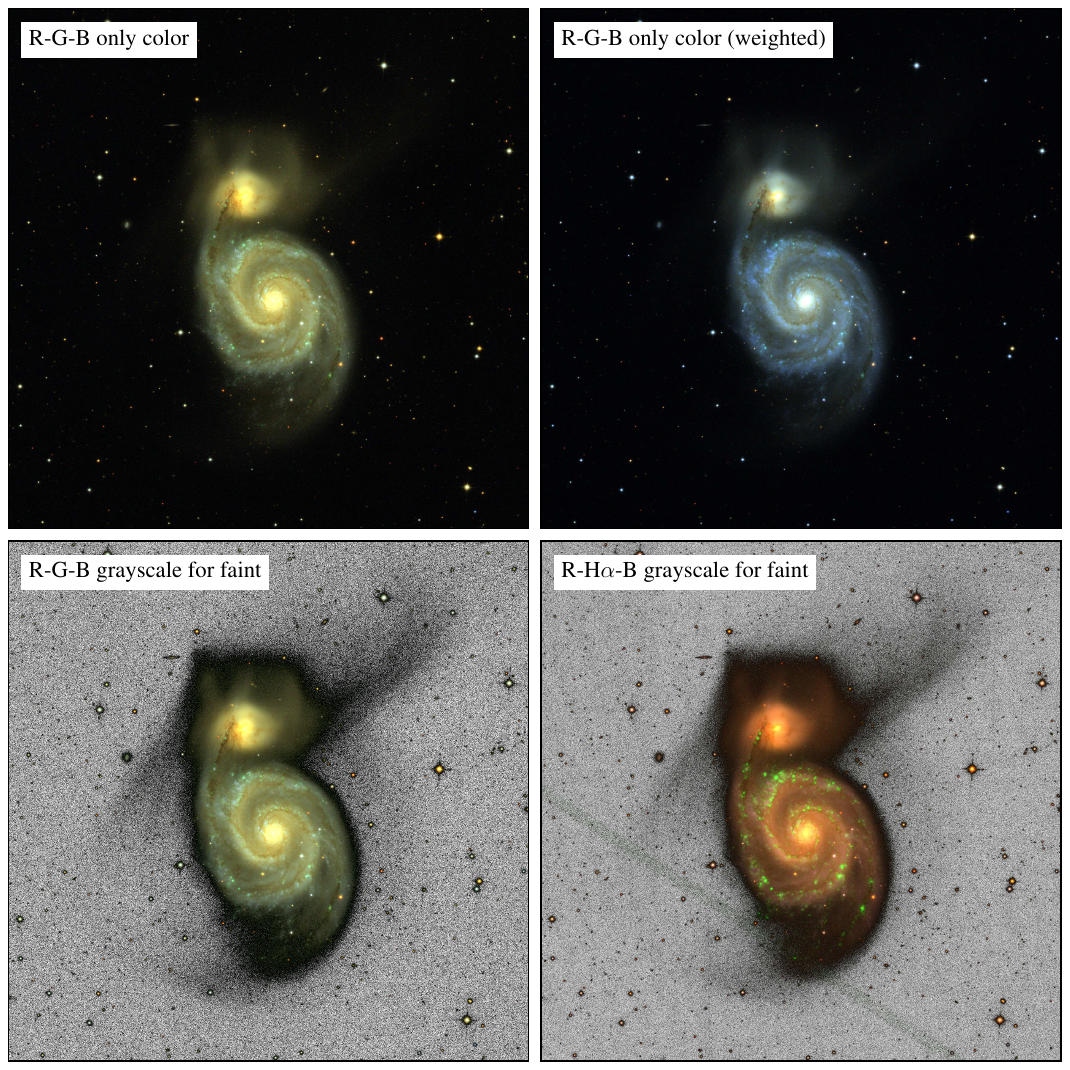}
  \fi

  \caption{
  \label{fig:m51color}
  \objname{} galaxy group color images from J-PLUS DR\jplusdr{} using the \chr, \chg, \chb{} (R-G-B), and \chh{} (H$\alpha$) filters.
  \emph{Top-left}: traditional color image, where the background regions become black.
  \emph{Top-right}: modified weights of the channels balance to obtain a bluer image.
  \emph{Bottom-left}: gray background color image; this is the default mode of \texttt{astscript-color-faint-gray}.
  The separation between color, black, and gray regions are defined from surface brightness cuts (see text) of the G channel (\sbch{}).
  The use of the gray background colormap reveals diffuse low surface brightness structures that would otherwise remain unveiled.
  \emph{Bottom-right}: color image using the H$\alpha$ narrow band filter (\chh) for the intermediate (G) channel instead of \chg.
  The use of this filter reveals interesting features such as the star-forming regions that are shown in green.
  }
\end{figure*}

In Figure~\ref{fig:m51color}, we present color images of the \objname{} galaxy group using J-PLUS \citep{cenarro2019} data.
Each image is obtained using different options.
Notably, the default gray background reveals intriguing low surface brightness features that are veiled, or saturated, in the traditional color image (that have a black background).
The different regions of the bottom-left panel correspond to a different surface brightness in the \sbch{} channel: colored regions are those brighter than $\sbbright\,\mast$, black regions are those regions between $\sbbright \text{ and }\sbfaint\,\mast$, while gray regions are those fainter than $\sbfaint\,\mast$.

The use of the H$\alpha$ narrow band filter (\chh) in the bottom-right panel allows for the visualization of interesting features such as the star-forming regions.
This shows how color images are not only useful for aesthetic/outreach images of astronomical sources but also to visualize quantitative measurements.

The noisy greenish line that is visible in the bottom-right panel shows that a good data reduction is important for this mode of displaying images: because fainter artifacts become visible with this colormap.
This exemplifies the power of fine-tuning the parameters to unveil valuable details in astronomical data.
Furthermore, due to the higher information content, the file size (in bytes) of gray images is larger than black images (this depends on the compression used in the output RGB format).
It is something to take into account when creating gray background images like this for the sky navigators of large astronomical surveys (e.g., Euclid, J-PAS, J-PLUS, Legacy Survey, LSST).

\section{Acknowledgments}
\label{sec:acknowledgments}
\noindent
The workflow uses Maneage \citep[\emph{Man}aging data lin\emph{eage},][commit \maneageversion]{maneage}.
This research note is created from the Git commit {\projectversion}, hosted on Codeberg\footnote{Git repository of the paper (\texttt{\projectgitbranch} branch): \url{\projectgitrepo}.} which is archived on Software Heritage\footnote{Software Heritage (SoftWare Hash IDentifier, SWHID): \href{https://archive.softwareheritage.org/swh:1:dir:1064a48d4bb58d6684c3df33c6633a04d4141d2d;origin=https://codeberg.org/gnuastro/papers;visit=swh:1:snp:a083ff647c571f895d1ccc9f7432fa1b9a1d03a8;anchor=swh:1:rev:ff77b619daa50b05ddd83206d979d1f8a53d040b}{swh:1:dir:1064a48d4bb58d6684c3df33c6633a04d4141d2d}} for longevity.
Supplements are also available on Zenodo\footnote{Zenodo: \url{https://doi.org/10.5281/zenodo.10058165}}.

The analysis uses GNU Astronomy Utilities \citep[Gnuastro,][]{gnuastro2015,gnuastro2019} v\gnuastroversion.
Gnuastro has been funded by the Japanese MEXT scholarship and its Grant-in-Aid for Scientific Research (21244012, 24253003), ERC 339659-MUSICOS, Spanish MINECO AYA2016-76219-P, and NextGenerationEU ICTS-MRR-2021-03-CEFCA.
We acknowledge the funding by Governments of Spain and Arag\'on through FITE and Science Ministry (PGC2018-097585-B-C21, PID2021-124918NA-C43).

%% Bibliography
\bibliography{references}{}
\bibliographystyle{aasjournal}

%% Appendix. Mention all used software.
\appendix
\section{Software acknowledgement}
\label{appendix:software}
 
This research was done with the following free software programs and libraries:  1.23, Bzip2 1.0.8, C compiler (Apple clang version 13.1.6 (clang-1316.0.21.2.5)), CFITSIO 4.1.0, CMake 3.24.0, Dash 0.5.11-057cd65, Discoteq flock 0.4.0, Expat 2.4.1, File 5.42, Fontconfig 2.14.0, FreeType 2.11.0, GNU AWK 5.1.1, GNU Astronomy Utilities 0.21.43-3101 \citep{gnuastro2015,gnuastro2019}, GNU Autoconf 2.71, GNU Automake 1.16.5, GNU Bash 5.2-rc2, GNU Bison 3.8.2, GNU Coreutils 9.1, GNU Diffutils 3.8, GNU Findutils 4.9.0, GNU Grep 3.7, GNU Gzip 1.12, GNU Libtool 2.4.7, GNU M4 1.4.19, GNU Make 4.3, GNU Multiple Precision Arithmetic Library 6.2.1, GNU Multiple Precision Floating-Point Reliably 4.1.0, GNU NCURSES 6.3, GNU Nano 6.4, GNU Readline 8.2-rc2, GNU Scientific Library 2.7, GNU Sed 4.8, GNU Tar 1.34, GNU Texinfo 6.8, GNU Wget 1.21.2, GNU Which 2.21, GNU gettext 0.21, GNU gperf 3.1, GNU libiconv 1.17, GNU libunistring 1.0, GPL Ghostscript 9.56.1, Git 2.37.1, Help2man , Less 590, Libffi 3.4.2, Libgit2 1.3.0, Libidn 1.38, Libjpeg 9e, Libpaper 1.1.28, Libpng 1.6.37, Libtiff 4.4.0, Libxml2 2.9.12, Lzip 1.23, OpenSSL 3.0.5, Perl 5.36.0, Python 3.10.6, WCSLIB 7.11, X11 library 1.8, XCB-proto (Xorg) 1.15, XZ Utils 5.2.5, Zlib 1.2.11, cURL 7.84.0, libICE 1.0.10, libSM 1.2.3, libXau (Xorg) 1.0.9, libXdmcp (Xorg) 1.1.3, libXext 1.3.4, libXt 1.2.1, libpthread-stubs (Xorg) 0.4, libxcb (Xorg) 1.15, pkg-config 0.29.2, podlators 4.14, util-Linux 2.38.1, util-macros (Xorg) 1.19.3, xorgproto 2022.1 and xtrans (Xorg) 1.4.0. 
The \LaTeX{} source of the paper was compiled to make the PDF using the following packages: courier 61719 (revision), epsf 2.7.4, etoolbox 2.5k, helvetic 61719 (revision), lineno 5.3, pgf 3.1.10, pgfplots 1.18.1, revtex4-1 4.1s, tex 3.141592653, textcase 1.04 and ulem 53365 (revision). 
We are very grateful to all their creators for freely  providing this necessary infrastructure. This research  (and many other projects) would not be possible without  them.

%% Finish LaTeX
\end{document}